\documentclass[a4paper,11pt]{article}
\usepackage{pos}

\usepackage[symbol]{footmisc}
\usepackage[english]{babel}
\usepackage{graphicx}
\usepackage{graphics}
\usepackage{braket}
\usepackage{bbold}
\usepackage{amsmath}
\usepackage{nicefrac}
\usepackage{dcolumn}
\usepackage{bm}
\usepackage{slashed}
\usepackage{datetime}
\usepackage{mciteplus}
\usepackage{multirow}
\usepackage{siunitx}
\usepackage{booktabs}
\usepackage{color, soul}
\usepackage[usenames,dvipsnames]{xcolor}
\usepackage{float}
\usepackage[utf8]{inputenc}
\usepackage[normalem]{ulem}
\usepackage{mathtools}
\usepackage{setspace}

\renewcommand{\thefootnote}{\fnsymbol{footnote}}

\newcommand{{\HFNRevo}}{\tt HF-NRevo}

\title{Towards Precision Gluon Densities at Small $x$: \\ From Resummation to Collider Observables}
\ShortTitle{Towards Precision Gluon Densities at Small $x$}

\author*[a]{Francesco Giovanni Celiberto}
\author[b]{Marco Bonvini}

\affiliation[a]{Universidad de Alcalá (UAH), Departamento de Física y Matemáticas, Campus Universitario, \\ Alcalá de Henares, E-28805, Madrid, Spain}

\affiliation[b]{INFN, Sezione di Roma 1, Piazzale Aldo Moro 5, I-00185 Roma, Italy}

\emailAdd{francesco.celiberto@uah.es}
\emailAdd{marco.bonvini@roma1.infn.it}

\abstract{Accurate gluon densities at small $x$ are essential for reducing theoretical uncertainties in collider predictions, yet remain one of the least constrained ingredients in global analyses. 
We report recent advances in the resummation of small-$x$ logarithms in the gluon sector, focusing on collinear distributions and their interplay with transverse-momentum-dependent formulations. Particular attention is paid to the impact of gluon-proton spin correlations and to the emergence of unintegrated gluon densities derived from resummed dynamics. 
These developments aim to fill the current precision gap in the small-$x$ region and enable robust applications to LHC and future-collider observables.}

\FullConference{The European Physical Society Conference on High Energy Physics (EPS-HEP2025)\\
7-11 July 2025\\
Marseille, France\\}

\begin{document}
\maketitle

\section{Hors d'{\oe}uvre}
\label{sec:introduction}

The precise determination of parton distribution functions (PDFs)~\cite{Ball:2017otu,Abdolmaleki:2018jln,Bonvini:2019wxf} at small values of the Bjorken variable $x$ is a central objective in modern hadron collider phenomenology. 
Gluon densities, in particular, dominate most high-energy processes, including (multi-)jet~\cite{Celiberto:2015yba,Caporale:2015int} production, heavy-flavour channels~\cite{Bolognino:2021mrc,Celiberto:2022dyf,Celiberto:2025euy,Silvetti:2022hyc,Celiberto:2024omj,Celiberto:2025ipt,Celiberto:2025dfe,Celiberto:2025ziy,Celiberto:2025vra}, Higgs boson dynamics~\cite{Hentschinski:2020tbi,Celiberto:2020tmb,Celiberto:2022fgx,Celiberto:2024bbv,DelDuca:2025vux}, and (forward) Drell--Yan processes~\cite{Celiberto:2018muu,Golec-Biernat:2018kem}. 
Despite their relevance, gluon PDFs remain among the least constrained components in global analyses, especially in the small-$x$ regime, where experimental coverage is limited and theoretical uncertainties are enhanced.

At small $x$, the growth of logarithmic terms of the type $\ln(1/x)$ severely hampers the reliability of fixed-order calculations. These logarithms originate from large rapidity separations and multiple soft gluon emissions, and their systematic resummation~\cite{Fadin:1975cb,Balitsky:1978ic} is essential to restore perturbative stability. 
Over the past two decades, considerable theoretical effort has been devoted to the inclusion of small-$x$ resummation within the collinear factorisation framework, allowing for more accurate PDF fits and cross-section predictions~\cite{Bonvini:2016wki,Bonvini:2017ogt,Bonvini:2018ixe,Bonvini:2018iwt}. 
Such improvements are particularly important in the context of the LHC and of next-generation colliders, where processes initiated by small-$x$ gluons will play a leading role.

Beyond inclusive observables, precision in the small-$x$ domain also requires a better grasp of less inclusive quantities that depend on the transverse dynamics of partons. The transition from traditional one-dimensional (1D) PDFs to three-dimensional (3D) distributions, such as transverse-momentum-dependent PDFs (TMD PDFs)~\cite{Bacchetta:2020vty,Celiberto:2021zww,Bacchetta:2024fci} and unintegrated gluon densities (UGDs)~\cite{Anikin:2011sa,Hentschinski:2012kr,Hentschinski:2013id,Bolognino:2018rhb,Celiberto:2019slj,Bolognino:2019pba,Bolognino:2021niq,Hentschinski:2021lsh}, enables a richer description of hadronic structure. 
These formulations retain information on partonic intrinsic motion, angular momentum, and polarisation, and are sensitive to spin-orbit correlations that become increasingly relevant at low $x$.

Moreover, the small-$x$ region is where novel QCD phenomena, such as parton saturation and nonlinear dynamics, are expected to emerge. A refined understanding of gluon dynamics in this regime is thus not only a matter of precision but also of principle, offering insights into the fundamental mechanisms of colour coherence and gluon recombination in dense systems.

In this contribution, we summarise recent advances in the construction of small-$x$ resummed gluon distributions, with emphasis on their connection to transverse-momentum-sensitive formulations. 
In particular, we highlight new strategies to extract UGDs from resummed collinear inputs, and we outline the perspectives for their use in high-energy processes at the LHC and beyond.

\section{From 1D to 3D tomography of the proton}
\label{sec:HASsx}

The most effective theoretical framework for probing the internal structure of the proton is the language of parton correlators. 
This formalism encodes the spatial, momentum, and energy distributions of the proton's elementary constituents---quarks and gluons, collectively known as partons. 
Among these tools, TMD PDFs stand out for their ability to provide a genuine 3D imaging of the proton. 
A TMD-based description is essential to address fundamental questions such as the origin of the proton spin~\cite{EuropeanMuon:1987isl}.

TMD PDFs generalise standard 1D collinear PDFs by incorporating the intrinsic transverse momentum of the struck parton. 
They are essential to describe observables sensitive to the transverse recoil of the hard probe, typically observed in semi-inclusive processes where the final-state particles exhibit transverse momenta or imbalances much smaller than the hard scale. TMD PDFs thus bridge the gap between internal proton dynamics and experimentally accessible quantities.

The complete set of twist-2 gluon TMD PDFs for a spin-1/2 target includes both unpolarised and polarised structures, following the naming conventions of Refs.~\cite{Meissner:2007rx,Lorce:2013pza}. 
The diagonal entries in this basis are $f_1^g$, the unpolarised gluon density in an unpolarised proton, and $g_1^g$, the distribution of circularly polarised gluons in a longitudinally polarised proton. In the collinear limit, they reduce to the well-known unpolarised and helicity gluon PDFs.

The distribution of linearly polarised gluons in an unpolarised hadron, $h_1^{\perp g}$, is particularly relevant because it leads to observable spin effects even in collisions of unpolarised hadrons with an expected enhancement in the small-$x$ region. 
TMD PDFs differ from their collinear counterparts in being process-dependent due to the presence of gauge links (Wilson lines) in their operator definitions~\cite{Ji:2002aa}, introducing additional theoretical complexity.

Although much is known about the perturbative resummation of transverse-momentum logarithms~\cite{Bozzi:2003jy,Catani:2010pd}, the genuinely nonperturbative content of gluon TMD PDFs remains largely unknown. 
Their experimental exploration is still in its infancy, especially in the gluon sector. 
The future Electron-Ion Collider (EIC)~\cite{AbdulKhalek:2021gbh,Amoroso:2022eow} is expected to open a new era in this field, but current efforts rely on flexible models to guide theory and phenomenology. 
These models provide essential insight and offer a basis for constructing realistic TMD PDFs that incorporate known constraints and anticipate small-$x$ dynamics.

A key step towards modeling the small-$x$ behavior of gluon TMD PDFs was achieved by adopting a spectator-inspired framework, where the struck gluon is embedded in a parent proton and its remnant is modeled as an effective on-shell particle~\cite{Bacchetta:2020vty,Celiberto:2021zww}.
This approach allows for a clean and flexible parameterization of spin-dependent TMD PDFs, through dipolar form factors that regulate the transverse-momentum structure and mitigate endpoint divergences. The inclusion of a spectral function over the spectator mass plays a crucial role in shaping the $x$-dependence of the distributions, enabling a smooth interpolation between the small- and moderate-$x$ regimes. 
All leading-twist, T-even gluon TMD PDFs were computed in this formalism and connected to their collinear counterparts through transverse-momentum integration.

Building upon this foundation, a subsequent step was taken to incorporate high-energy dynamics via small-$x$ resummation.
Collinear gluon PDFs, both unpolarised and helicity, were obtained using two distinct NLO inputs: the {\tt NNPDF3.1sx} set~\cite{Ball:2017otu}, which includes small-$x$ resummation through the HELL formalism~\cite{Bonvini:2016wki,Bonvini:2017ogt,Bonvini:2018ixe,Bonvini:2018iwt}, and the {\tt NNPDFpol1.1} set~\cite{Nocera:2014gqa}, for the helicity channel. 
These 1D distributions were used at the lowest common hadronic scale, before the respective evolution schemes (DGLAP and TMD) cause a kinematic decorrelation between the collinear and transverse sectors.

The unpolarised and helicity gluon TMDs were then fitted simultaneously to their collinear counterparts through transverse-momentum integration. This fit procedure allowed for a consistent determination of the spectral-mass profile and the dipolar form factors regulating the nonperturbative content.
Once the parameters were fixed, all T-even gluon TMD PDFs at leading twist were generated within the same spectral framework, resulting in the first tomographic reconstruction of the proton at small $x$ based on resummed inputs.
Work in Ref.~\cite{Bacchetta:2024fci} extended this study to the T-odd sector.
These results capture key features of 3D gluon dynamics at small $x$, combining theoretical consistency with flexibility, and offering a realistic basis for guiding future phenomenology and data-driven extraction strategies in the pre-EIC era.

\section{Towards precision UGDs at small $x$}
\label{sec:UGDs}

The UGDs provide a powerful and versatile framework for describing the internal structure of the proton in the small-$x$ regime.
These two-dimensional (2D) distributions, depending on both the longitudinal momentum fraction $x$ and the transverse momentum $|\vec k_T|$ of the gluon, serve as a bridge between fully collinear PDFs and more elaborate, resummation-based or non-linear QCD descriptions of hadronic dynamics at high energies.

We remark that UGDs and TMD PDFs are distinct objects, as they are defined within different factorisation frameworks---the high-energy-factorisation approach in the former case, and the TMD factorisation scheme in the latter.
As a consequence, they obey different evolution equations and encode complementary aspects of QCD dynamics, particularly in the small-$x$ and moderate-$x$ regimes respectively.

In contrast to their 1D collinear counterparts, UGDs retain an explicit dependence on the gluon transverse momentum, thereby capturing key features of transverse dynamics that are crucial for forward and semi-inclusive processes. 
Their relevance spans a wide array of physical contexts, ranging from high-energy factorisation approaches to gluon saturation phenomena, and they represent essential building blocks in calculations performed within the $k_T$-factorisation formalism and the colour-glass-condensate (CGC) effective theory~\cite{Dominguez:2011wm}.

Historically, the extraction of UGDs has been challenged by the limited theoretical control over relevant observables beyond leading logarithmic (LL) accuracy, and by the scarcity of direct experimental constraints in the small-$x$ regime. 
Although the conceptual foundations of UGDs are well established, their practical implementation in high-energy QCD analyses often relies on simplified assumptions or phenomenological models, due to the absence of comprehensive NLL-level descriptions for many processes sensitive to the $(x,k_T)$ dynamics.

As a result, most UGDs currently available in the literature are obtained via indirect procedures, typically by tuning model parameters to reproduce transverse-momentum spectra of forward particles~\cite{Anikin:2011sa,Hentschinski:2012kr,Hentschinski:2013id,Bolognino:2018rhb,Bolognino:2019pba,Bolognino:2021niq}. 
These approaches, while valuable for exploratory studies and qualitative understanding, fall short of the rigour required for precision phenomenology. In particular, they usually neglect a systematic treatment of theoretical uncertainties and are not guaranteed to be consistent with collinear PDFs extracted from global fits.
All this has hindered the development of a unified and quantitatively reliable framework for describing gluon dynamics at small $x$, particularly in applications requiring a precise control over both longitudinal and transverse degrees of freedom.

This has led to a fundamental issue: despite their central role in small-$x$ physics, UGDs remain among the least constrained and most poorly understood parton distributions. 
The gap between their theoretical significance and practical determination has become increasingly evident in light of high-luminosity experimental programmes and the growing need for accurate predictions in the small-$x$ domain.

To address this issue, we have initiated the development of a new strategy for constructing precision UGDs directly linked to resummed collinear inputs. The central idea is to enforce a consistent matching between the UGD and its integrated counterpart: by construction, integration over transverse momentum must yield the collinear gluon PDF at a fixed factorisation scale. 
This requirement establishes a firm and theoretically controlled link between the two- and 1D descriptions of the proton's gluon content, forming the basis for a systematic, data-informed construction of UGDs.

A crucial element of our strategy is the adoption of collinear gluon distributions that incorporate small-$x$ resummation at next-to-leading logarithmic (NLL) accuracy, implemented through the HELL formalism. This approach builds upon the foundational principles established by the Altarelli–Ball–Forte (ABF) approach~\cite{Ball:1995vc,Altarelli:2003hk,Altarelli:2005ni}, ensuring a consistent treatment of high-energy logarithms and the stabilization of the high-energy resummation kernel. 
As a result, the collinear input used in our construction is both theoretically robust and phenomenologically reliable. The inclusion of resummation leads to a stabilisation of the small-$x$ behaviour, avoiding unphysical enhancements and yielding controlled, realistic evolution down to very low values of $x$.

In this preliminary stage of development, these resummed collinear inputs are embedded into a flexible parametrisation of the unintegrated gluon density, shaped by spectral functions that regulate the transverse-momentum profile and ensure smooth integrability. The construction is designed so that the integration over transverse momentum faithfully recovers the input collinear gluon distribution within estimated uncertainties, which can then be propagated consistently to the two-dimensional UGD level.

This methodology is currently under development, and early results are promising. For the first time, we are able to generate UGDs that include rigorous uncertainty estimates, grounded in resummed collinear fits and suitable for immediate application to high-energy QCD observables.

The resulting UGDs will be interfaced with the JETHAD~\cite{Celiberto:2020wpk,Celiberto:2022rfj} framework, enabling direct applications to forward observables at the LHC and future colliders. This includes studies of decorrelation patterns, heavy-flavour production, and gluon-sensitive observables where a detailed treatment of the $(x,|\vec k_T|)$ dependence is essential. Through JETHAD, we aim to systematically explore the contributions of different kinematic regions to measurable cross sections and to assess how resummation-induced effects shape forward distributions.

\section{Summary and outlook}
\label{sec:conclusions}

We have outlined a unified research strategy aimed at bridging complementary approaches to the proton’s internal structure in the small-$x$ regime. 
Starting from 1D collinear PDFs enhanced by small-$x$ resummation, we developed consistent methods to construct both 3D TMD distributions and 2D UGDs. 
Each of these frameworks offers a distinct window onto gluon dynamics, from inclusive observables to angular correlations and forward spectra, and is grounded in robust theoretical inputs and controlled uncertainty estimates. Throughout this programme, emphasis has been placed on maintaining compatibility with global fits, leveraging high-energy factorisation tools, and embedding the entire construction within a numerically efficient and phenomenologically oriented environment.

The long-term vision is to realise a triangular strategy: one that simultaneously connects 1D collinear PDFs, 3D TMD distributions, and 2D UGDs within a common, data-informed framework. By correlating the nonperturbative parameters across these three descriptions and fitting them to different classes of observables, we aim to achieve a multidimensional understanding of proton structure at small $x$. 
This integrated approach---combining rigour, flexibility, and high phenomenological reach---marks a decisive step toward precision QCD in the high-energy limit and opens the path to a qualitatively new generation of fits.

\section*{Acknowledgments}
\label{sec:acknowledgments}

F.G.C. is supported by the Atracción de Talento Grant no. 2022-T1/TIC-24176 of the Comunidad Autónoma de Madrid, Spain, and by the INFN/QFT@COLLIDERS Project, Italy.
MB is supported by the Italian Ministry of University and Research (MUR) grant PRIN 2022SNA23K funded by the European Union -- Next Generation EU, Mission 4, Component 2, CUP I53D23001410006.

\begingroup
\setstretch{0.6}
\bibliographystyle{bibstyle}
\bibliography{biblography}
\endgroup

\end{document}